# Coherent Relation between Structure and Conduction of Infinite Atomic Wires

Tomoya Ono[1], Hideki Yamasaki[2], Yoshiyuki Egami[2] and Kikuji Hirose[2]

[1]Research Center for Ultra-Precision Science and Technology, Osaka University, Suita, Osaka 565-0871, Japan;

[2]Department of Precision Science and Technology, Osaka University, Suita, Osaka 565-0871, Japan.

**Abstract**

We demonstrate a theoretical analysis concerning the geometrical structures and electrical conduction of infinite monatomic gold and aluminum wires in the process of their elongation, based on first-principles molecular-dynamics simulations using the real-space finite-difference method. Our study predicts that the single-row gold wire ruptures up to form a dimer coupling structure when the average interatomic distance increases up to more than 3.0 Å, and that the wire is conductive before breaking but changes to an insulator at the rupturing point. In the case of the aluminum wire, it exhibits a magnetic ordering due to the spin polarization, and even when stretched up to the average interatomic distance of 3.5 Å, a dimerization does not occur and the wire keeps a metallic nature.

## 1. Introduction

In the last several years, the properties of metallic wire contacts have attracted great attention, and many experiments concerning wire contacts have been carried out using a scanning tunneling microscope and a mechanically controllable break junction[1]. The simultaneous measurement of mechanical force and conductance and also the direct observation by *in situ* transmission electron microscopy gave us visible information about the relationship between geometrical structure and electrical conduction of the wire[1-5]. In this nanometer-scale fabrication, the conductance through the wire having a diameter of the order of the electron Fermi wavelength is quantized in an unit of $G_0 = 2e^2/h$ (*e*: the electron charge, *h*: Planck's constant), when the interatomic distance is gradually varied. So far, a large number of theoretical and empirical studies concerning the wires have been reported, however, there still remain some questions: how long the maximum interatomic distance of the wire is and how much the conductance of the single-row wire is.

In this paper, we present a theoretical analysis of the geometrical structures and electron conduction of infinite monatomic wires on the basis of first-principles molecular-dynamics (FPMD) simulations within the framework of the density functional theory[6]. We found that when a single-row gold wire is elongated gradually, the conductance of the wire is quantized at the unit of $G_0$, and finally at the average interatomic distance of 3.0 Å, the Peierls distortion occurs to form a dimer coupling structure inside the wire. On the other hand, a single-row aluminum wire exhibits a magnetic ordering due to spin-polarized electronic ground state, and no dimerization occurs even when the average interatomic distance increases up to 3.5 Å, much larger than the nearest neighbor spacing of the bulk crystal (2.2 Å).

The organization of the present paper is as follows: In Sec. 2 we describe briefly the method used in our calculation. Our results are presented and discussed in Sec. 3. We summarize our findings in Sec. 4.

## 2. Computational method

We employ here FPMD simulation program based on the real-space finite-difference (RSFD) method[7]. The RSFD method can determine ground-state electronic and atomic structures to a high degree of accuracy with a modest overhead in the computational cost, thanks to the timesaving double-grid technique[8] and the direct minimization of the energy functional[9]. This method has the following advantages: First, it does not need any basis-function set, unlike linear-combination-atomic-orbitals method in which one suffers from the difficulties that theoretical results depend on the choice of basis sets and there is no straightforward way to prepare a complete basis. Second, the standard linear-combination-atomic-orbitals method fails to describe accurate tails of the wave functions. It should be noticed that for a single-row wire where the atoms might be spaced by 3.0–4.0 Å, these tails in the interstitial region between adjacent atoms have a particularly significant meaning. The RSFD method is a neat approach to enable us to take the tails into account accurately. Another advantage to be noticed is that the RSFD method tackles serious drawbacks of the plane-wave approach, e.g., its inability to describe strictly nonperiodic systems such as clusters and solid surfaces.

Figure 1 depicts the calculation model, where two atoms are included in a supercell of $12.7 \times 12.7 \times 2d_{av}$ Å$^3$. Here $d_{av}$ is the average interatomic distance, i.e., the wire length per atom. The boundary condition is periodic in the $z$ direction which is parallel to the wire. In the $x$ and $y$ directions, we impose the nonperiodic boundary condition of vanishing wave function out of the supercell, in order to eliminate completely unfavorable effects of atoms in neighbor cells which are artificially repeated in the case of the periodic boundary condition. As an initial configuration for FPMD, the parameters of the atomic geometry $d_1$ and $\theta$ at $d_{av}$=2.3 Å are randomly set in the ranges of 2.3±0.1 Å and 160±10 degree, respectively, and these are optimized through FPMD calculations. Then, we stretch the wire and relieve the force on atoms repetitiously.

We obey the nine-point finite-difference formula for the derivative arising from the kinetic-energy operator, and the dense-grid spacing is fixed at $h_{dens}=h_{coars}/3$, where $h_{coars}$ is the coarse grid spacing[8]. The norm conserving scalar relativistic pseudopotentials including partial core corrections are employed in a Kleinman-Bylander nonlocal form[10,11]. Exchange-correlation effects are treated with the local-spin-density approximation (LSDA)[12] and the generalized gradient approximation (GGA)[13] in which spin degrees of freedom are taken into account. The Brillouin-zone integrations are performed using the Monkhorst and Pack 6 $k_z$-point prescription[14]. We take the cutoff energy, which is determined as $(\pi/h_{coars})^2$, to be 82 and 25 Ry for the gold and aluminum wires, respectively.

## 3. Results and discussion

Figure 2 shows the cohesive energy, the restoring force, and the parameters in the fully optimized geometry of the single-row wire as a function of the average intera-tomic distance $d_{av}$. Closed circles and crosses represent the results in the cases of the LSDA and GGA, respectively. For the gold wire, there is no evidence that the wire in the electronic ground state is spin-polarized all over the range of $d_{av}$ in both the cases of the LSDA and GGA calculations. The maximum restoring force of the gold wire is 2.0 (1.3) nN in the LSDA (GGA) calculation. These values are similar to that observed empirically by Rubio *et al*.[2] The gold wire breaks when pulled by the maximum restoring force, i.e., beyond the maximum average interatomic distance of 3.0 Å. This theoretical result prefers the experimental data obtained by Kizuka *et al*.[5], the critical distance at the fracture of 3.2 Å, to the data by Ohnishi *et al*.[3] of 3.5–4.0 Å and by Yanson *et al*.[4] of 3.6 Å. After the breaking, two atoms form a dimer coupling on a single straight line, i.e., the Peierls dimerization occurs, as seen in Figs. 2 (c) and (d)[15]. The spacing of atoms in the dimer is 2.55 Å, which is equal to the equivalent bond length of $Au_2$[16].

On the other hand, the ground-state electronic structure of the infinite aluminum wire is spin-polarized and the tension of the wire is about 1.1 nN [Fig.2 (g)], which is of the same magnitude for the case of gold wire. When the wire is stretched up to $d_{av}$=2.9 Å, the spin-polarized electronic ground state sensibly emerges. This result is in agreement with the other theoretical indication[17]. It is interesting that the wire consisting solely of nonmagnetic element alone shows the magnetic property. What is more, elongated up to the average atomic distance of 3.5 Å, a value much larger than the equivalent bond length of $Al_2$ (2.5 Å) and the nearest neighbor spacing of bulk crystal (2.2 Å), the aluminum wire does not yield a dimer coupling structure. In the previous study, the electronic ground state of $Al_2$ molecule was found to be in $^3\Sigma_g^-$ state[16], which is a spin-polarized configuration. We thus explore a possibility that the elongated aluminum wire forms a dimer coupling structure in antiferromagnetic ordering. The wire at $d_{av}$=2.96 Å is taken as example for this purpose and we employ a supercell of $12.7 \times 12.7 \times 4d_{av}$ Å$^3$ to replicate the dimer coupling structure in antiferromagnetic ordering and perform the Brillouin-zone integration over 30 $k_z$ points[14]. Atoms B and D in the wire are moved simultaneously toward the others along the wire axis to from dimers with atoms A and C, respectively (see inset in Fig.

3). The cohesive energy as a function of the bond length of the dimmer, $d_1$, is shown in the Fig. 3. One can note that the wire in the antiferromagnetic ordering is slightly more stable than that in the ferromagnetic ordering, and the wire where the atoms are equi-spaced is the most stable atomic configuration in the both cases of the LSDA and GGA calculations. Therefore we can conclude that the wire where the atoms are equi-spaced in ferromagnetic ordering is the most stable atomic and electronic configuration.

We next discuss the electrical conduction of the single-row wires. Figs. 4 and 5 show the energy band structure and the partial density of state (PDOS) at various average interatomic distances in the LSDA calculation, respectively. Here, we employed 50 $k_z$-point prescription using symmetrized Monkhorst-Pack grids[14], to implement accurately the summation over the Brillouin zone. From Fig. 5 one can see that for the case of the gold wire, the conduction process is realized practically only through $s$-$p_z$ orbital, since $d$ bands are located deeply below the Fermi level, while for the case of the aluminum wire, doubly-degenerate $p_x$-$p_y$ orbitals contribute to conduction. According to the Hund rule, these partially filled $p_x$-$p_y$ symmetry bands give rise to the spin-polarized ground state of the aluminum wire.

Finally, we show in Fig. 6 the electronic conductances of the monatomic wires as a function of the average interatomic distance. The conductance $G$ is calculated by the linear-response Kubo formula[18],

$$G = \lim_{E \to E_F} \frac{\pi e^2 \hbar}{m^2 L_z^2} \sum_{\sigma=\uparrow\downarrow} \left| \langle \psi_\sigma(E) | \hat{p}_z | \psi_\sigma(E_F) \rangle \right|^2 [N_\sigma(E_F)]^2,$$

where $\{\psi_\sigma\}$ are single-particle wave functions with spin index $\sigma$ normalized in the supercell, $\hat{p}_z = -i\hbar\partial/\partial z$, $m$ is the electron mass, $L_z$ is the length of the supercell in $z$ direction and $N_\sigma(E_F)$ is the density of states relevant to the occupation of $\sigma$-spin electrons at the Fermi level $E_F$. Since only electrons at the Fermi level contribute to conductance, we are only interested in $\psi_\sigma(E_F)$. The single-row gold wire before breaking has a conductance of 1 $G_0$. When the average interatomic distance exceeds 3.0 Å, the conductance changes to zero, i.e., the gold wire transfers from a metal to an insulator at its rupturing point. This theoretical result is in correspondence with the experimen-tal data obtained by Ohnishi *et al.*[3] and by Yanson *et al.*[4] that the conductance is quantized as 1 $G_0$ right before the rupturing point, and disagrees with the data of zero conductance by Kizuka *et al.*[5] On the other hand, the conductance of the aluminum wire at $d_{av}$=2.54 Å is 2 $G_0$. When we stretch the wire, the conductance changes to 3 $G_0$ at $d_{av}$≈2.8 Å. After the wire exhibits the maximum conductance of 3 $G_0$, the conductance decreases in the unit of $G_0$ with the increase of the average interatomic distance. Thus the value of the conductance for the infinite aluminum wire is at variance with the experimental and theoretical results[19,20] yielding only 1 $G_0$ for the finite aluminum wire. The discrepancy may be explained by the interaction between the wire and electrodes. Moreover, at $d_{av}$≈3.5 Å, the conductance is 2 $G_0$, i.e., the wire remains conductive. It is amazing that the wire with the atoms spaced by such a long distance as $d_{av}$≈3.5 Å exhibits a metallic nature. However, we explored only the possibility of distortion with the dimerized structure, and if the aluminum wire broke and formed the other atomic geometry, e.g., the structure including trimer or tetramer molecules, at a short average interatomic distance, then such an elongated wire might be artificial.

## 4. Conclusion

The coherent relation between the atomic structure and electronic conduction of infinite single-row wires was established by FPMD calculations. In the case of the gold wire, our result, which predicts 3.0 Å as the maximum stretch of a bond for the single chain of gold atom, is in accordance with the experimental data obtained by Kizuka *et al.*[5] On the other hand, the conductance is about 1 $G_0$ before the wire breaks, and when stretched over the rupturing point, the wire transfers from a metal to an insulator. This result supports the experiments by Ohnishi *et al.*[3] and Yanson *et al.*[4] In the case of the aluminum wire, the maximum conductance of the infinite single-row wire is 3 $G_0$. The fracture of the wire and the metal-insulator transition are not observed even when the average interatomic distance increases up to more than 3.5 Å. It is unexpected that the atoms are spaced by such a long distance in the infinite single-row wire and the conductance of the wire keeps 2 $G_0$. Here, we have stud-ied only the distortion mechanism with the dimerized structure, therefore we must study another type of the mechanism, such as trimarization or tetramerization, in a future study.

**Acknowledgments**

This research was partially supported by the Ministry of Education, Culture, Sports, Science and Technology, Grant-in-Aid for Young Scientists (B), 14750022, 2002. The numerical calculation was carried out by the computer facilities at the Institute for Solid State Physics at the University of Tokyo.

# Figure captions

Fig. 1 Model for an infinite single-row wire adopted in the calculation. The size of supercell is taken as $L_x=L_y=12.7$ Å and $L_z=2d_{av}$, where $d_{av}$ is the average interatomic distance. The closed circles denote the atoms in a supercell and the open ones show their replicated atoms in the adjacent supercells.

Fig. 2 Calculated results for the cohesive energy [(a) and (f)], the restoring force [(b) and (g)], the bond length $d_1$ [(c) and (h)], the bond length $d_2$ [(d) and (i)], and the bond angle θ [(e) and (j)] in the infinite single-row wires. Circles (crosses) show the results obtained by the LSDA (GGA) calculation. The geometric parameters of the structure ($d_1$, $d_2$, θ) are explained in Fig. 1.

Fig. 3 Cohesive energy as a function of bond length of dimer in the infinite aluminum wire with ferromagnetic (FM) and antiferromagnetic (AFM) ordering. The calculated points are fit to spline-interpolated curves as a guide to eye. The broken vertical line indicates the position where the atoms are equi-spaced by 2.96 Å in the wire. Inset: Atomic geometry of dimerized wires. The length of the supercell along the wire axis $4d_{av}$ is taken to be 11.84 Å. Atom C locates at the center between A and A'. Atoms B and D are replaced simultaneously, and A and C are fixed at the same position.

Fig. 4 Energy band structure at various average interatomic distances $d_{av}$, plotted along the direction $k_z$ in reciprocal space. The solid (broken) curves represent the contributions of the up- (down-) spin electrons. The zero of energy is chosen to be the Fermi level.

Fig. 5 Partial density of states (PDOS) distribution at various average interatomic distances $d_{av}$. The solid, dotted, and broken curves represent the PDOS of the $s$-$p_z$, $p_x$-$p_y$, and $d$ levels, respectively. The zero of energy is chosen to be the Fermi level.

Fig. 6 Calculated results for the conductance versus the average interatomic distance $d_{av}$. Circles (triangles) show the conductance of the infinite gold (aluminum) wire.

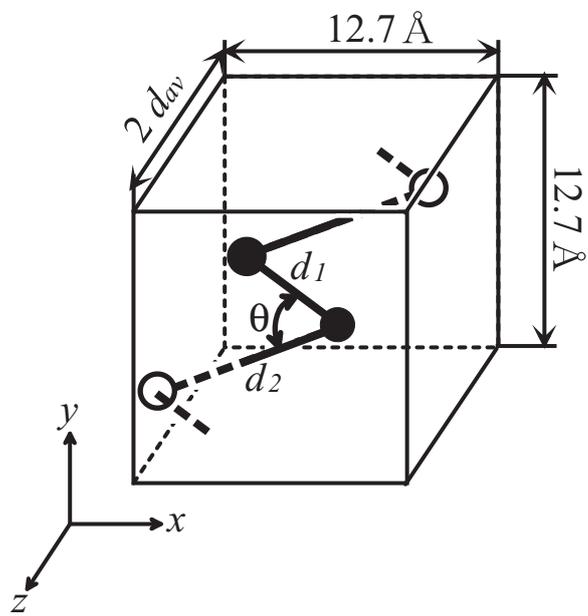

Fig. 1

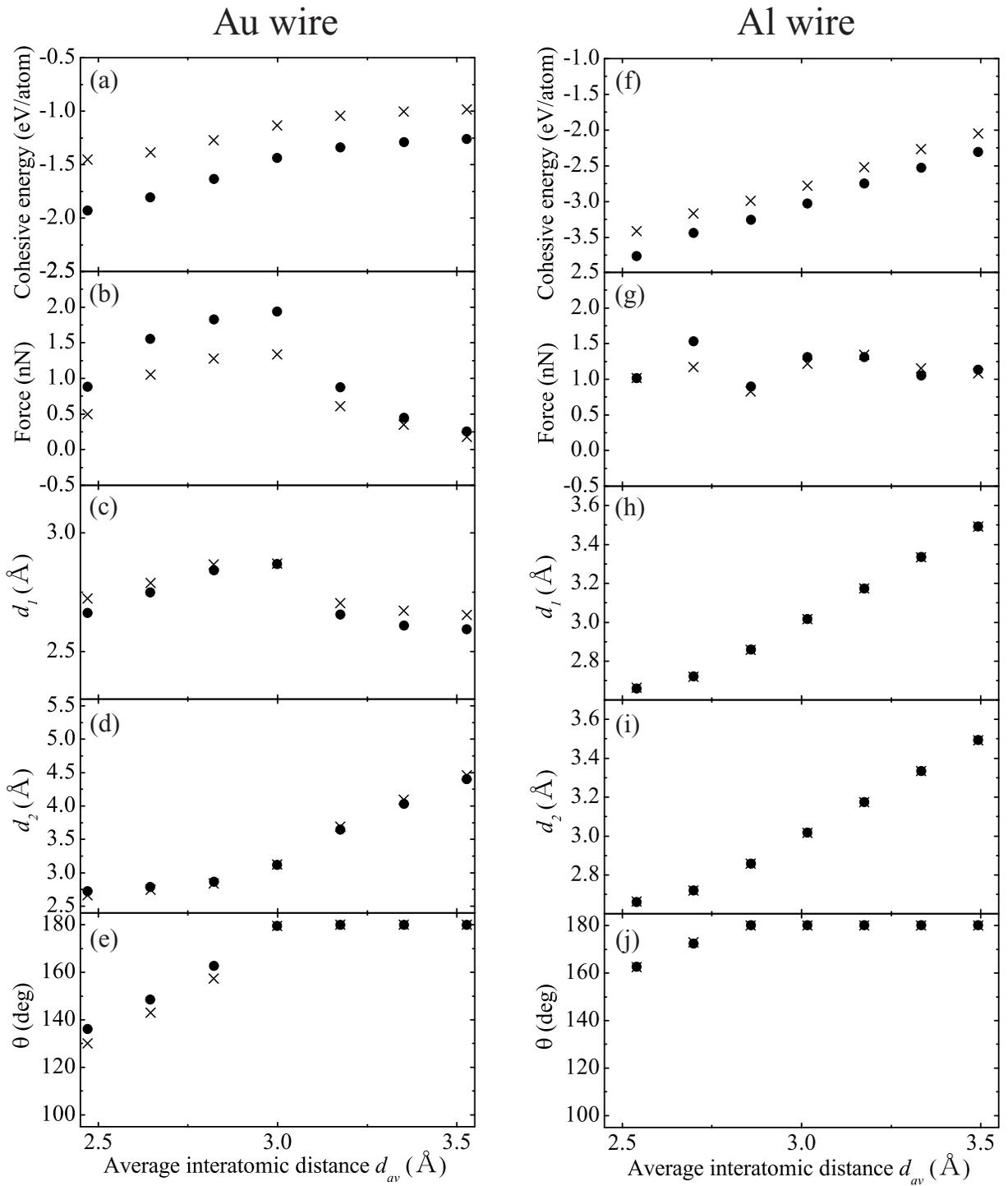

Fig. 2

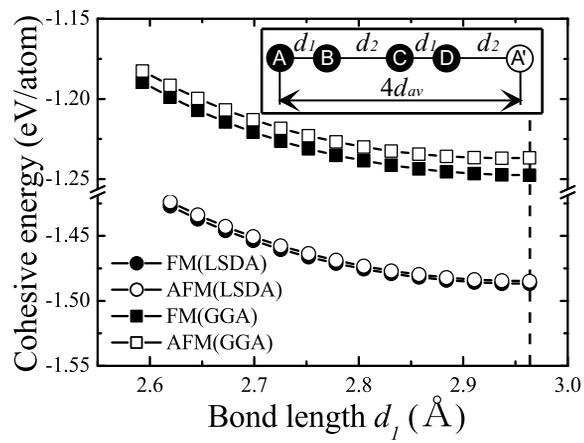

Fig. 3

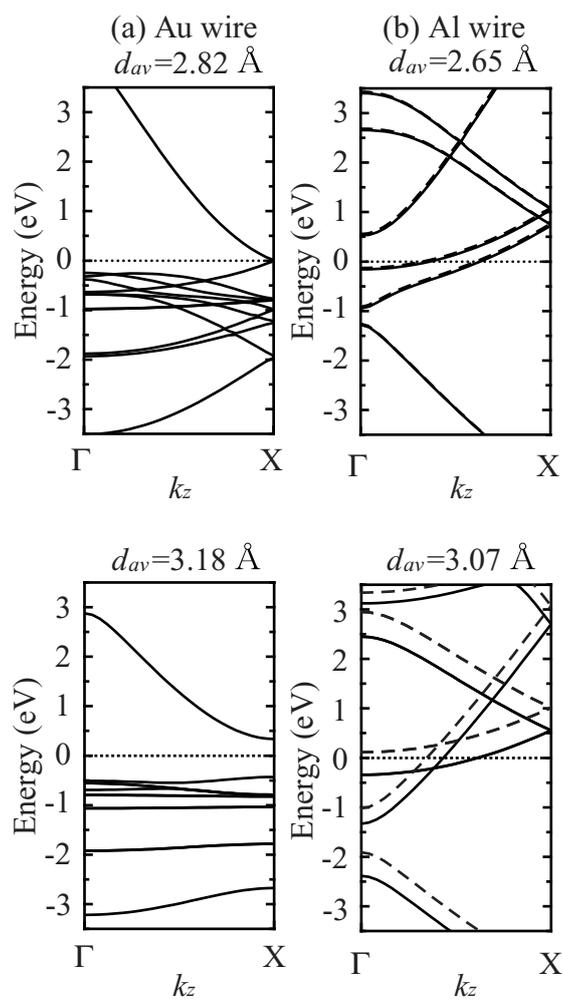

Fig. 4

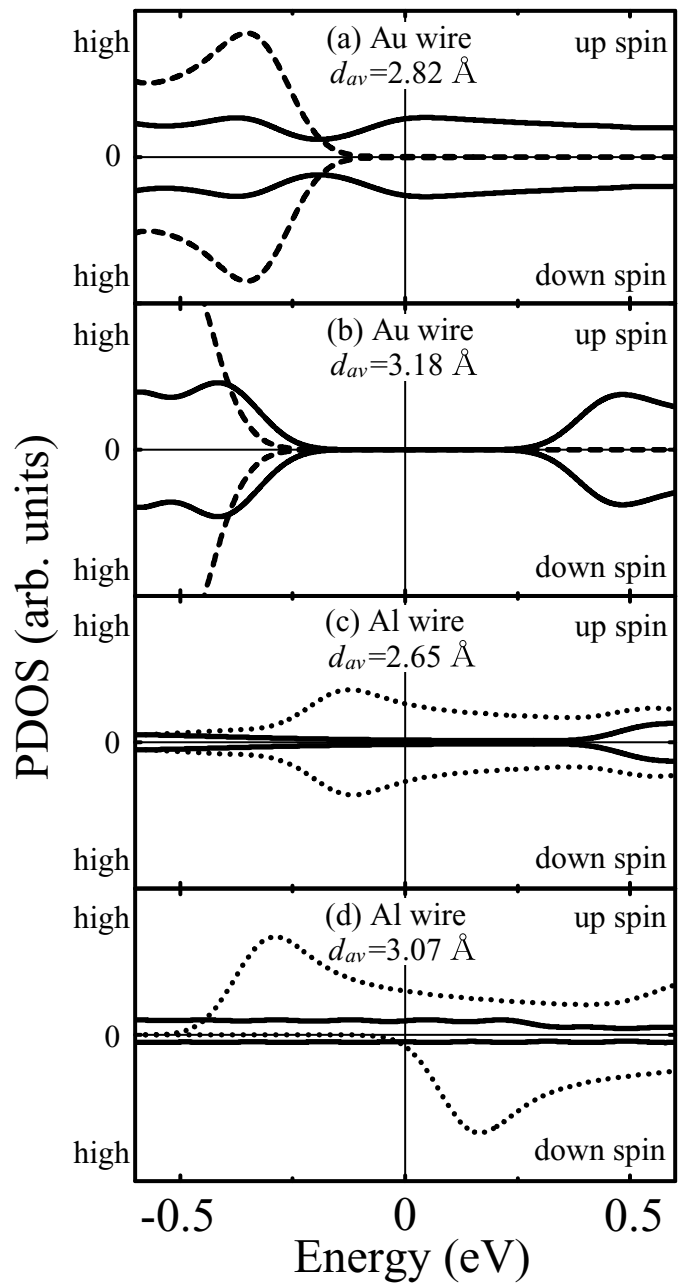

Fig. 5

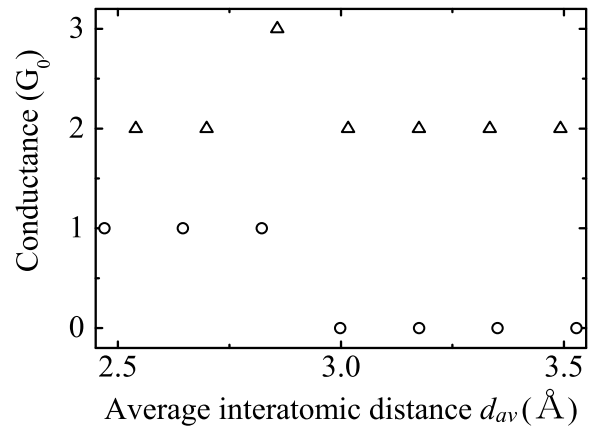

Fig. 6